\documentclass[aip,apl,reprint]{revtex4-1}
\usepackage{units}
\usepackage{color}
\usepackage{amstext}
\usepackage{amssymb}
\usepackage{amsmath}
\usepackage{graphicx}

\begin{document}

\title{Undoped accumulation-mode Si/SiGe quantum dots}

\author{M.~G.~Borselli}
\email{mborselli@hrl.com}
\author{K.~Eng}
\author{R.~S.~Ross}
\author{T.~M.~Hazard}
\author{K.~S.~Holabird}
\author{B.~Huang}
\author{A.~A.~Kiselev}
\author{P.~W.~Deelman}
\author{L.~.D.~Warren}
\author{I.~Milosavljevic}
\author{A.~E.~Schmitz}
\author{M.~Sokolich}
\author{M.~F.~Gyure}
\author{A.~T.~Hunter}
\noaffiliation

\affiliation{HRL Laboratories,\,LLC, Malibu, CA 90265, USA}

\date{15 July 2014}

\begin{abstract}

We report on a quantum dot device design that combines the low disorder properties of undoped SiGe heterostructure materials with an overlapping gate stack in which each electrostatic gate has a dominant and unique function --- control of individual quantum dot occupancies and of lateral tunneling into and between dots. Control of the tunneling rate between a dot and an electron bath is demonstrated over more than nine orders of magnitude and independently confirmed by direct measurement within the bandwidth of our amplifiers. The inter-dot tunnel coupling at the $(0,2)\leftrightarrow(1,1)$ charge configuration anti-crossing is directly measured to quantify the control of a single inter-dot tunnel barrier gate.  A simple exponential dependence is sufficient to describe each of these tunneling processes as a function of the controlling gate voltage.

\end{abstract}

\maketitle

%[Introduction material]

Silicon-based quantum devices hold great promise for realizing spin qubits.\cite{Zwanenburg2013} The ability to isotopically purify silicon has resulted in the demonstration of extremely long spin-coherence times in donor-based silicon devices\cite{Zwanenburg2013,Muhonen2014} and in recent years a series of results have demonstrated many fundamental properties of electrostatically-defined silicon-based quantum devices. Measurements of $\text{T}_1$,\cite{Hayes2009,*Xiao2010PRL,*Simmons2011} valley splitting,\cite{Shi2011,Prance2012} and Pauli blockade\cite{Shaji2008,Prance2012} were made using doped depletion-mode SiGe devices. Improved device performance was achieved by eliminating the intentional dopants in the SiGe heterostructure, a major source of noise and instability, making necessary the use of a global field gate to accumulate electrons. This allowed for demonstrations of high mobility two-dimensional electron gases (2DEGs),\cite{Lu2009} Coulomb blockade,\cite{Lu2011} valley splitting,\cite{Xiao2010APL} Pauli blockade,\cite{Borselli_FG}and $\text{T}_2^\star$ and Rabi measurements.\cite{Maune2012}

The ideal realization of electrostatically-defined quantum dot devices would have independent  control of the charge occupancy of each dot and its associated exchange couplings. A promising approach is to utilize an accumulation-based design in which independent localized gates are used to create electron baths, create and control quantum dot occupancy, and modulate tunnel barriers between them. A Si metal-oxide-semiconductor based design has shown great promise along these lines, demonstrating in quick succession charge sensing,\cite{Yang2011,Yang2012} valley splitting,\cite{Lim2011} and well-controlled double-dot behavior including spin blockade.\cite{Lai2011}  However better isolation from residual disorder due to gate oxide charges can potentially be achieved using a SiGe heterostructure.

In this Letter we report on a double quantum dot device with an integrated dot charge sensor based on a synthesis of the improved gated control of the accumulation-mode designs\cite{Lai2011} with the lower disorder of field-gated SiGe heterostructure designs.\cite{Borselli_FG}  This approach builds on our previous experience with quantum devices made using single gates in accumulation mode\cite{Croke2010,Borselli_VS}  and achieves the goal of complete gated control over a set of quantum dots and inter-dot couplings, dominating over the effects of disorder.

% [Main text of paper]

\begin{figure}
\centering
\includegraphics[width=3.2in]{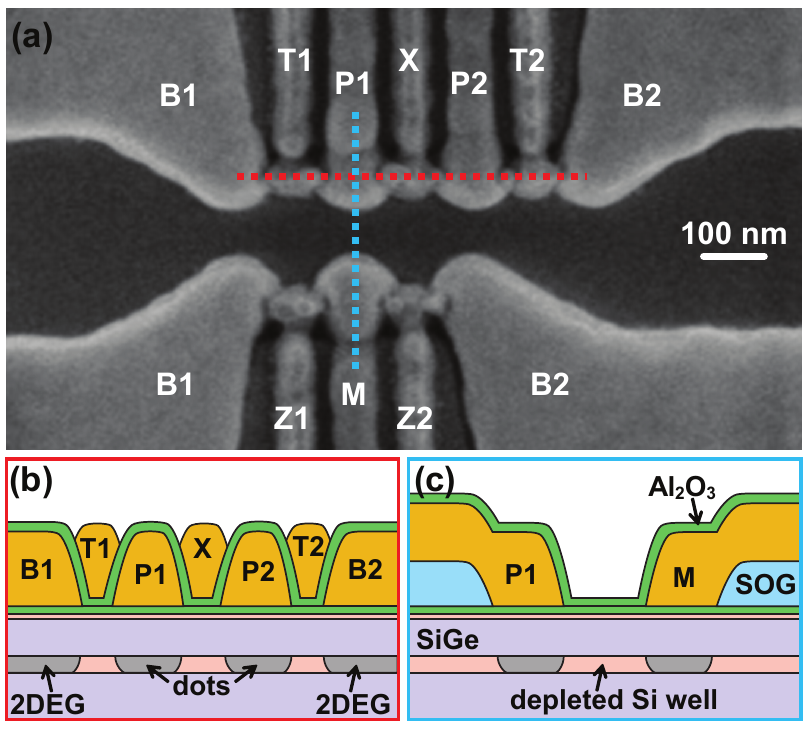}
\caption{\label{fig:Fig1} (a) Scanning electron micrograph of the double quantum dot device with adjacent dot charge sensor.  (b) Schematic cross-section through coupled double dots with adjacent 2DEG baths (shown as a red dashed line in (a)).  (c) Schematic cross-section through the dots formed by the P1 and M gates with depleted region in between (shown as vertical blue dashed line in (a)).}
\end{figure}

A nominally undoped epitaxial heterostructure similar to that previously reported by our group\cite{Borselli_FG} produces a tensily-strained Si quantum well on a strain-relaxed $\text{Si}_{0.7}\text{Ge}_{0.3}$ buffer.  Alternating depositions of dielectric and metal are patterned via electron-beam lithography to create an overlapping set of highly localized gates along the edge of, and adjacent to, a low-$k$ dielectric spin-on glass (SOG).\cite{Wolf_and_TauberV1}  Shown schematically in Fig.~\ref{fig:Fig1}, these gates can be operated in forward bias to locally control the creation of 2DEG baths (B1 and B2) and quantum dots (P1, P2, and M), and the intervening tunnel barriers (T1, X, T2, Z1, and Z2). Each of the non-bath gates have leads that enter the central device area on top of the SOG. The lower capacitance of these leads results in a higher threshold voltage, preventing charge accumulation in the quantum well except under the tips of the gates that touch down closer to the epitaxial layers.

\begin{figure}
\centering
\includegraphics[width=3.2in]{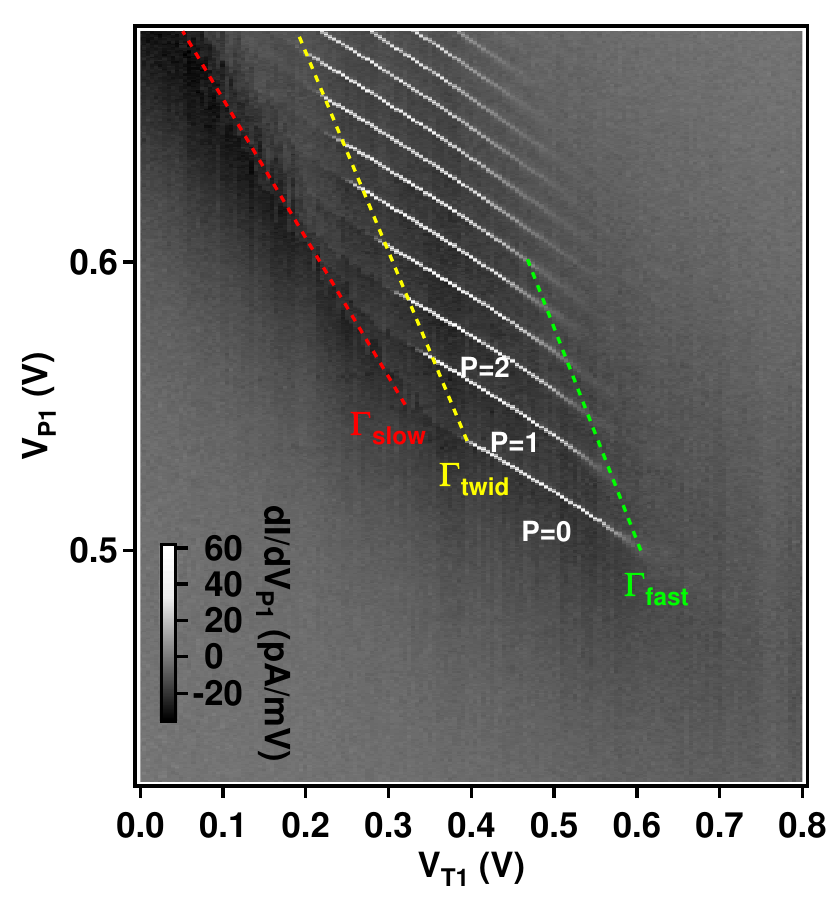}
\caption{\label{fig:Fig2} Charge stability diagram of a single dot connected to a 2DEG bath. Differential transconductance ($dI/dV_\text{P1}$) is mapped as a function of gate voltages on P1 and T1 as depicted in Fig.~\ref{fig:Fig1}. White lines in data indicate biases in which charge is exchanged between the dot and the reservoir at the pulse frequency of $388\,\text{Hz}$.  Dashed lines are guides to the eye indicating iso-tunneling rates comparable to: bias scan rate (red), pulse frequency (yellow), and lifetime broadening (green).}
\end{figure}

% [Forming dot-based charge sensor]
An integrated quantum dot charge sensor is formed by biasing the lower set of gates (B1, Z1, M, Z2, B2) such that a few-electron dot is created under gate M that is strongly tunnel coupled to 2DEGs formed under bath gates B1 and B2, with nominal electron densities $n_\text{2DEG}\approx5\times10^{11}~\text{cm}^{-2}$.  The voltage on M is first set to the few-electron regime (based on the threshold voltage of a large-area Hall bar) and the tunnel barrier gates, Z1 and Z2, are adjusted to create symmetric tunneling rates between the sensing dot (M-dot) and the 2DEGs.  The voltage on M is then set to the edge of a Coulomb blockade peak where $dI/dV_M$ is at a maximum.\cite{Barthel2010,Nordberg2009}  The charge sensitivity is determined by the ratio of the P- to M-dot mutual capacitance, $C_\text{PM}$, to the product of their total capacitances as ${\Delta I}/{I} \propto {C_\text{PM}} / ({C_\text{P}^\text{tot} C_\text{M}^\text{tot} } )$.\cite{Nordberg2009} These devices have a typical maximum sensitivity $\Delta I / I$ of $20\%$.

In order to probe the device electrostatics over a large range of bias space with constant sensitivity, the chemical potential of the M-dot is held approximately fixed. For small changes in the voltage on M, the gain is constant allowing measurement of the capacitive ratios of the P, T, and X gates to the chemical potential of the M-dot. These provide linear voltage corrections to $V_\text{M}$ which are used in a software-controlled open-loop feedback to straightforwardly hold fixed the chemical potential of the sensing dot as the other gate voltages are varied.

%[forming plunger dots]
Charge stability diagrams of quantum dots formed under the plunger gates, P1(2), are generated by sweeping the DC biases, $V_\text{P1(2)}$, versus a respective neighboring tunnel barrier gate voltage, $V_\text{T1(2)}$. Figure~\ref{fig:Fig2} shows an example (from a device loaded in a ${^3\text{He}}$\,--${^4\text{He}}$ dilution refrigerator with a base temperature of $20\,\text{mK}$) generated by applying a small-amplitude, $\delta =\pm 0.5\,\text{m\!V}$, square wave at $f_\text{pulse}=388\,\text{Hz}$ to the P1 gate and recording the differential transconductance with standard lock-in techniques. In the region between the red and yellow lines, the tunneling rate into the dot is slower than the gate pulse twiddle frequency,\cite{Elzerman2004} $\Gamma < \Gamma_\text{twid}=2\pi f_\text{pulse}=2.4\!\times\!10^3\,\text{s}^{-1}$, but faster than the sweep rate of $\sim\!72\,\text{s}$ per column of data. Between the yellow and green lines, the tunneling rate into the P1-dot becomes faster than the gate pulse twiddle frequency, $\Gamma > \Gamma_\text{twid}$, resulting in stable charge transition boundaries. For values of $V_\text{T1}$ to the right of the green line, the tunneling rate into the dot exceeds that associated with the peak-to-peak amplitude of the square pulse applied to P1, $\Gamma > \Gamma_\text{fast} \simeq 2\delta/\hbar\sim10^{12}\,\text{s}^{-1}$, resulting in lifetime broadening of the transition lines. The ratio of the rates is $\Gamma_\text{fast} / \Gamma_\text{twid}\simeq4\!\times\!10^8$, and furthermore, the observation of single shot events down to the scan rate, $\ll 1\text{Hz}$, means that the T1 gate is controlling the dot-to-bath tunnel rate by well over nine orders of magnitude.

The relative capacitances of the P1 and T1 gates to the quantum dot are not changing, i.e., the iso-occupancy lines are straight, indicating that the position of the centroid of the dot wavefunction in the 2DEG remains approximately fixed.  The absence of other transitions or inflections indicates a system with low disorder, and this stability allows the simple open-loop feedback control to maintain constant sensitivity for hours. An advantage of this accumulation-mode design over previous work\cite{Borselli_FG} is the marked decrease in cross-capacitance between adjacent gates. In depletion-mode designs it is not uncommon for several gates to have comparable capacitances to the dot chemical potential of interest.\cite{Hanson2007}  By accumulating charge under a single gate, as is done here, the capacitive ratios of the adjacent gates are significantly reduced.  The largest cross-capacitance term for any quantum dot in this device is from its adjacent tunnel barrier gate and is of order twenty percent (slope of charge transition boundaries $V_\text{P1} / V_\text{T1}$ in Fig.~\ref{fig:Fig2}).

\begin{figure}
\centering
\includegraphics[width=3.2in]{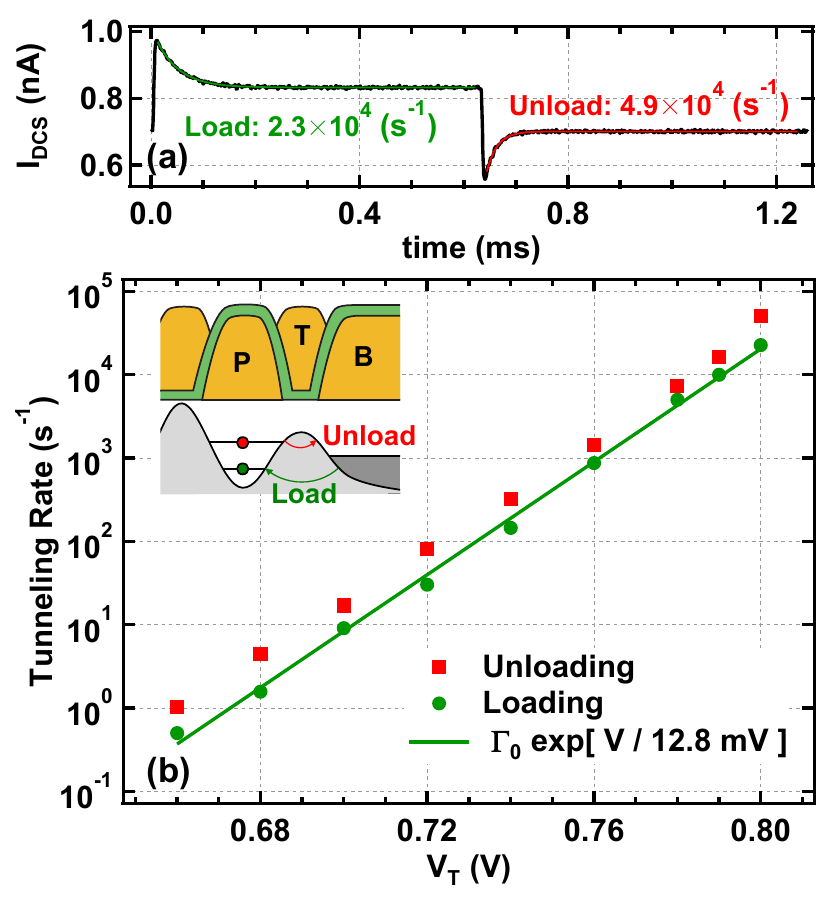}
\caption{\label{fig:Fig3} (a) Ensemble-averaged charge sensor current while an electron is repeatedly loaded and unloaded from a bath, as schematically depicted in inset to (b). The charge sensitivity is $\sim\!\!180\,\text{pA}/e^-$. (b) Plot of repeated measurements of tunneling rates as depicted in (a) as a function of the T-gate voltage.  The solid green line is a fit to the loading rate data illustrating the simple exponential tunneling rate dependence versus gate bias.}
\end{figure}

%[Mapping tunneling rates in time domain]
A quantitative analysis of the dot-to-bath tunneling rates was obtained from a second similar device  loaded in a $^3$He refrigerator with a base temperature of $270\,\text{mK}$ and tuned to a comparable configuration. The loading and unloading rates of the first electron as a function of the tunnel barrier gate bias, $V_\text{T}$, were measured by applying a fixed $\pm 2\,\text{m\!V}$ amplitude square pulse to the plunger gate.
\footnote{Independent excited-state-spectroscopy measurements\cite{Elzerman2004} and magneto-spectroscopy\cite{Borselli_VS,Yang2012,Shi2011} confirmed that this pulse amplitude was large enough to indiscriminately load into both ground and excited valley states.}
The pulse durations were varied to be approximately one order of magnitude longer than the measured rates. For each value of $V_\text{T}$, both $V_\text{P}$ and $V_\text{M}$ were linearly compensated to  hold the chemical potentials of the P- and M-dots approximately fixed with respect to the 2DEG. The charge sensor currents were recorded (averaged over 7,800 time traces in Fig.~\ref{fig:Fig3}(a)) and fit to exponential time decays to extract the average loading and unloading rates. The resulting tunneling rates, plotted in Fig.~\ref{fig:Fig3}(b) as a function of T gate voltage, have a simple exponential dependence over nearly five orders of magnitude. Simulations of these device designs confirm that, under the conditions of these experiments, the dominant influence of the T gate is to modulate the strength of the potential barrier between the dot and 2DEG, resulting in the observed exponential dependence on gate voltage. The unloading rate is always faster than the loading rate over the entire range, as is expected due to the smaller tunnel barrier in the unloading configuration.\cite{Thalakulam2011}

\begin{figure}
\centering
\includegraphics[width=3.2in]{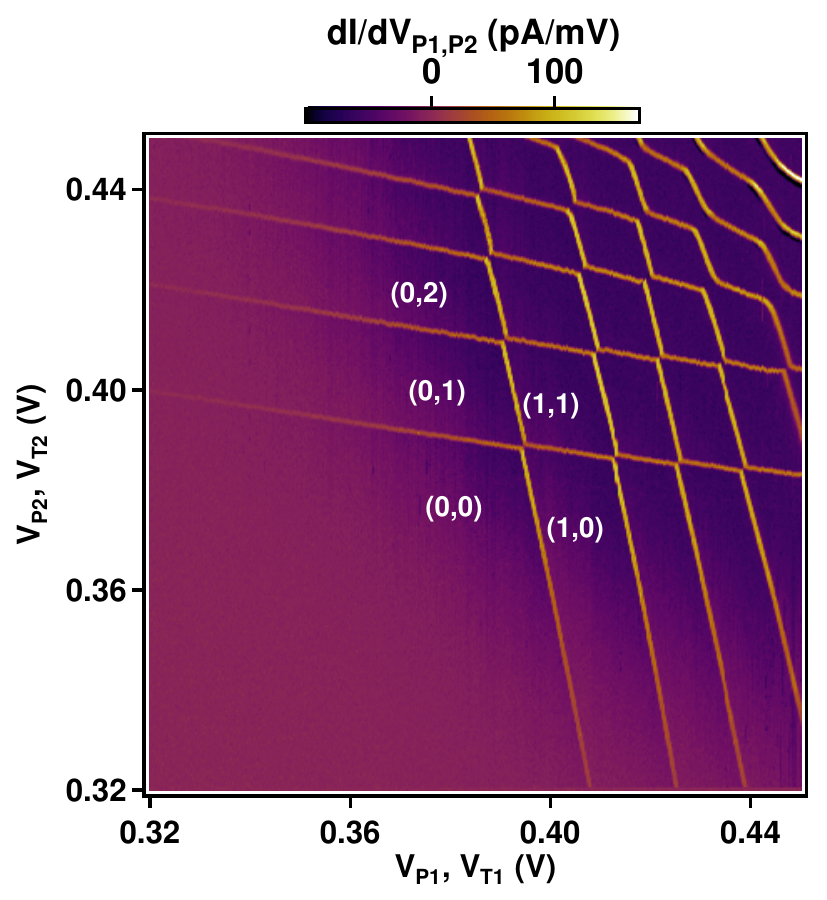}
\caption{\label{fig:Fig4} Charge stability diagram in $V_\text{P1}$ vs $V_\text{P2}$ taken in a dilution refrigerator. Linear capacitive corrections to $V_\text{M}$ ensure nearly constant charge sensor gain, and to $V_\text{T1}$, $V_\text{T2}$ maintain nearly constant, and open, dot-to-bath tunnel barriers. The inter-dot coupling gate $X$ was fixed; $V_\text{X}=0.8\,\text{V}$.  A $40\,\mu\!\text{V}$ DC bias was applied between the charge sensor source and drain, and a $\pm 0.35\,\text{m\!V}$ square wave amplitude was applied to both $V_\text{P1}$ and $V_\text{P2}$. }
\end{figure}

%[Forming a coupled double dot]
Figure~\ref{fig:Fig4} shows a charge stability diagram as a function of the voltages on gates P1 and P2, demonstrating the formation of a well-behaved double quantum dot. The dot-to-bath tunneling rates have been held approximately fixed by linearly compensating with voltages $V_\text{T1}$ and $V_\text{T2}$. The inter-dot tunnel coupling is being allowed to change as we are holding the X-gate voltage fixed. The observed increase in tunnel coupling towards the upper right of the figure is due to the small cross-capacitance of the P1(2) gates to the inter-dot potential barrier; therefore increasing positive bias on the P gates lowers the tunnel barrier with respect to the Fermi level. We are confident in the attributed charge configurations as no evidence of further transitions is seen along the respective first electron loading lines while maintaining barrier transparency.

\begin{figure}
\centering
\includegraphics[width=3.2in]{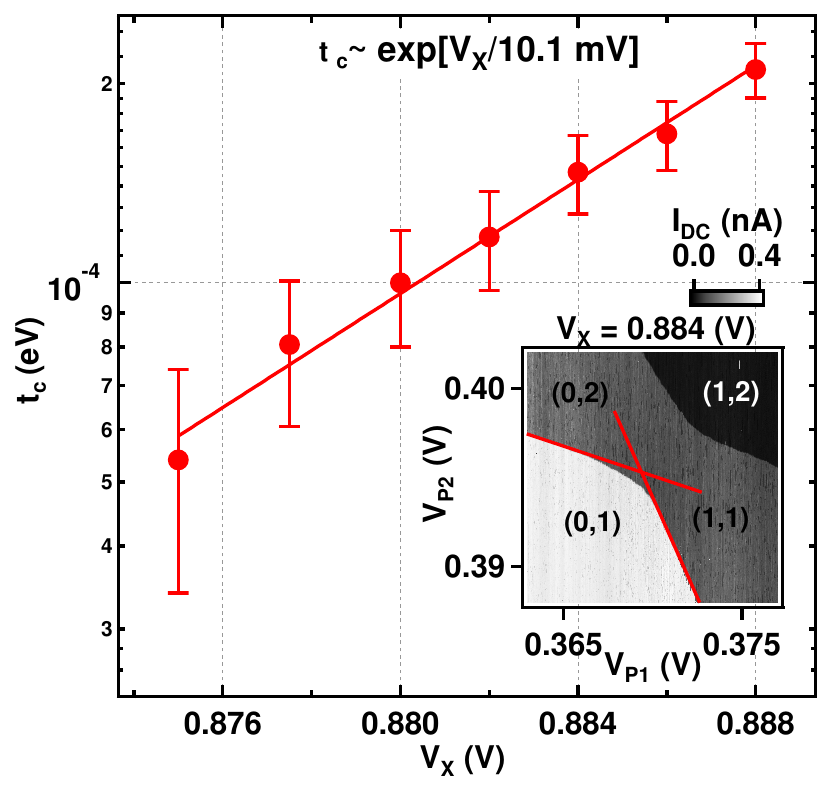}
\caption{\label{fig:Fig5} Inter-dot tunnel coupling, $t_c$, measured from the curvature of charge transition lines in the neighborhood of the $(0,2)\leftrightarrow(1,1)$ triple points.\cite{Hanson2007} (inset) Example charge stability diagram for $V_\text{X}=0.884~\text{V}$ from which curvature is determined with asymptotic charge boundaries shown in red.}
\end{figure}

%[Quantifying exchange coupling via DC transport]
Finally we demonstrate control of the inter-dot tunnel coupling at the $(0,2)\leftrightarrow(1,1)$ charge transition. Larger values of tunnel coupling can be measured directly from the curvature of charge state boundaries at the triple points of a charge stability diagram.\cite{Hanson2007}  In Fig.~\ref{fig:Fig5} we plot the tunnel coupling extracted from $V_\text{P1}$ versus $V_\text{P2}$ charge stability diagrams like that shown in the inset. For each fixed X-gate voltage we map the charge sensor current while sweeping the P1 and P2 gates. The charge sensor voltage, $V_\text{M}$, is compensated to maximize sensitivity to changes in total charge resulting in no observable differential charge sensitivity between the $(0,2)$ and $(1,1)$ charge configurations. This technique maximizes the contrast of the net-charge boundaries allowing measurement of $t_c$ down to $\sim 50\,\mu\text{eV}$.  The tunnel coupling is highly controllable and follows a similar simple exponential dependence on the controlling gate voltage as was found for the dot-to-bath tunnel coupling. A similar exponential dependence of the inter-dot tunnel coupling on a controlling gate voltage has been reported previously in a SiGe depletion-mode double quantum dot.\cite{Simmons2009}

%[Summary]
We have demonstrated the basic operation of SiGe quantum dot devices that use overlapping gates to individually control electron occupancy and lateral tunneling. The electron tunneling rates between quantum dots and adjacent electron baths, and between pairs of quantum dots exhibit a simple exponential dependence on a single controlling gate voltage. The variation of the dot-to-bath tunneling rate has been shown to be controllable over at least nine orders of magnitude. This device design, utilizing an undoped SiGe heterostructure and operated in an all-accumulation-mode, achieves robust gated electrostatic control of few-electron systems in an inherently low-disorder environment.

%\begin{acknowledgments}
Sponsored by United States Department of Defense. The views and conclusions contained in this document are those of the authors and should not be interpreted as representing the official policies, either expressly or implied, of the United States Department of Defense or the U.S. Government. Approved for public release, distribution unlimited.
%\end{acknowledgments}

%\bibliography{SiQuantumDots}

\begin{thebibliography}{27}%
\makeatletter
\providecommand \@ifxundefined [1]{%
 \@ifx{#1\undefined}
}%
\providecommand \@ifnum [1]{%
 \ifnum #1\expandafter \@firstoftwo
 \else \expandafter \@secondoftwo
 \fi
}%
\providecommand \@ifx [1]{%
 \ifx #1\expandafter \@firstoftwo
 \else \expandafter \@secondoftwo
 \fi
}%
\providecommand \natexlab [1]{#1}%
\providecommand \enquote  [1]{``#1''}%
\providecommand \bibnamefont  [1]{#1}%
\providecommand \bibfnamefont [1]{#1}%
\providecommand \citenamefont [1]{#1}%
\providecommand \href@noop [0]{\@secondoftwo}%
\providecommand \href [0]{\begingroup \@sanitize@url \@href}%
\providecommand \@href[1]{\@@startlink{#1}\@@href}%
\providecommand \@@href[1]{\endgroup#1\@@endlink}%
\providecommand \@sanitize@url [0]{\catcode `\\12\catcode `\$12\catcode
  `\&12\catcode `\#12\catcode `\^12\catcode `\_12\catcode `\%12\relax}%
\providecommand \@@startlink[1]{}%
\providecommand \@@endlink[0]{}%
\providecommand \url  [0]{\begingroup\@sanitize@url \@url }%
\providecommand \@url [1]{\endgroup\@href {#1}{\urlprefix }}%
\providecommand \urlprefix  [0]{URL }%
\providecommand \Eprint [0]{\href }%
\providecommand \doibase [0]{http://dx.doi.org/}%
\providecommand \selectlanguage [0]{\@gobble}%
\providecommand \bibinfo  [0]{\@secondoftwo}%
\providecommand \bibfield  [0]{\@secondoftwo}%
\providecommand \translation [1]{[#1]}%
\providecommand \BibitemOpen [0]{}%
\providecommand \bibitemStop [0]{}%
\providecommand \bibitemNoStop [0]{.\EOS\space}%
\providecommand \EOS [0]{\spacefactor3000\relax}%
\providecommand \BibitemShut  [1]{\csname bibitem#1\endcsname}%
\let\auto@bib@innerbib\@empty
%</preamble>
\bibitem [{\citenamefont {Zwanenburg}\ \emph {et~al.}(2013)\citenamefont
  {Zwanenburg}, \citenamefont {Dzurak}, \citenamefont {Morello}, \citenamefont
  {Simmons}, \citenamefont {Hollenberg}, \citenamefont {Klimeck}, \citenamefont
  {Rogge}, \citenamefont {Coppersmith},\ and\ \citenamefont
  {Eriksson}}]{Zwanenburg2013}%
  \BibitemOpen
  \bibfield  {author} {\bibinfo {author} {\bibfnamefont {F.~A.}\ \bibnamefont
  {Zwanenburg}}, \bibinfo {author} {\bibfnamefont {A.~S.}\ \bibnamefont
  {Dzurak}}, \bibinfo {author} {\bibfnamefont {A.}~\bibnamefont {Morello}},
  \bibinfo {author} {\bibfnamefont {M.~Y.}\ \bibnamefont {Simmons}}, \bibinfo
  {author} {\bibfnamefont {L.~C.~L.}\ \bibnamefont {Hollenberg}}, \bibinfo
  {author} {\bibfnamefont {G.}~\bibnamefont {Klimeck}}, \bibinfo {author}
  {\bibfnamefont {S.}~\bibnamefont {Rogge}}, \bibinfo {author} {\bibfnamefont
  {S.~N.}\ \bibnamefont {Coppersmith}}, \ and\ \bibinfo {author} {\bibfnamefont
  {M.~A.}\ \bibnamefont {Eriksson}},\ }\href@noop {} {\bibfield  {journal}
  {\bibinfo  {journal} {Rev. Mod. Phys.}\ }\textbf {\bibinfo {volume} {85}},\
  \bibinfo {pages} {961} (\bibinfo {year} {2013})}\BibitemShut {NoStop}%
\bibitem [{\citenamefont {Muhonen}\ \emph {et~al.}(2014)\citenamefont {Muhonen}
  \emph {et~al.}}]{Muhonen2014}%
  \BibitemOpen
  \bibfield  {author} {\bibinfo {author} {\bibfnamefont {J.~T.}\ \bibnamefont
  {Muhonen}} \emph {et~al.},\ }\href@noop {} {} (\bibinfo {year} {2014}),\
  \Eprint {http://arxiv.org/abs/1402.7140v1} {arXiv:1402.7140v1 [cond-mat]}
  \BibitemShut {NoStop}%
\bibitem [{\citenamefont {Hayes}\ \emph {et~al.}(2009)\citenamefont {Hayes}
  \emph {et~al.}}]{Hayes2009}%
  \BibitemOpen
  \bibfield  {author} {\bibinfo {author} {\bibfnamefont {R.}~\bibnamefont
  {Hayes}} \emph {et~al.},\ }\href@noop {} {} (\bibinfo {year} {2009}),\
  \Eprint {http://arxiv.org/abs/0908.0173v1} {arXiv:0908.0173v1 [cond-mat]}
  \BibitemShut {NoStop}%
\bibitem [{\citenamefont {Xiao}, \citenamefont {House},\ and\ \citenamefont
  {Jiang}(2010{\natexlab{a}})}]{Xiao2010PRL}%
  \BibitemOpen
  \bibfield  {author} {\bibinfo {author} {\bibfnamefont {M.}~\bibnamefont
  {Xiao}}, \bibinfo {author} {\bibfnamefont {M.~G.}\ \bibnamefont {House}}, \
  and\ \bibinfo {author} {\bibfnamefont {H.~W.}\ \bibnamefont {Jiang}},\
  }\href@noop {} {\bibfield  {journal} {\bibinfo  {journal} {Phys. Rev. Lett.}\
  }\textbf {\bibinfo {volume} {104}},\ \bibinfo {pages} {096801} (\bibinfo
  {year} {2010}{\natexlab{a}})}\BibitemShut {NoStop}%
\bibitem [{\citenamefont {Simmons}\ \emph {et~al.}(2011)\citenamefont
  {Simmons}, \citenamefont {Prance}, \citenamefont {Bael}, \citenamefont {Koh},
  \citenamefont {Shi}, \citenamefont {Savage}, \citenamefont {Lagally},
  \citenamefont {Joynt}, \citenamefont {Friesen}, \citenamefont {Coppersmith},\
  and\ \citenamefont {Eriksson}}]{Simmons2011}%
  \BibitemOpen
  \bibfield  {author} {\bibinfo {author} {\bibfnamefont {C.~B.}\ \bibnamefont
  {Simmons}}, \bibinfo {author} {\bibfnamefont {J.~R.}\ \bibnamefont {Prance}},
  \bibinfo {author} {\bibfnamefont {B.~J.~V.}\ \bibnamefont {Bael}}, \bibinfo
  {author} {\bibfnamefont {T.~S.}\ \bibnamefont {Koh}}, \bibinfo {author}
  {\bibfnamefont {Z.}~\bibnamefont {Shi}}, \bibinfo {author} {\bibfnamefont
  {D.~E.}\ \bibnamefont {Savage}}, \bibinfo {author} {\bibfnamefont {M.~G.}\
  \bibnamefont {Lagally}}, \bibinfo {author} {\bibfnamefont {R.}~\bibnamefont
  {Joynt}}, \bibinfo {author} {\bibfnamefont {M.}~\bibnamefont {Friesen}},
  \bibinfo {author} {\bibfnamefont {S.~N.}\ \bibnamefont {Coppersmith}}, \ and\
  \bibinfo {author} {\bibfnamefont {M.~A.}\ \bibnamefont {Eriksson}},\
  }\href@noop {} {\bibfield  {journal} {\bibinfo  {journal} {Phys. Rev. Lett.}\
  }\textbf {\bibinfo {volume} {106}},\ \bibinfo {pages} {156804} (\bibinfo
  {year} {2011})}\BibitemShut {NoStop}%
\bibitem [{\citenamefont {Shi}\ \emph {et~al.}(2011)\citenamefont {Shi},
  \citenamefont {Simmons}, \citenamefont {Prance}, \citenamefont {Gamble},
  \citenamefont {Friesen}, \citenamefont {Savage}, \citenamefont {Lagally},
  \citenamefont {Coppersmith},\ and\ \citenamefont {Eriksson}}]{Shi2011}%
  \BibitemOpen
  \bibfield  {author} {\bibinfo {author} {\bibfnamefont {Z.}~\bibnamefont
  {Shi}}, \bibinfo {author} {\bibfnamefont {C.~B.}\ \bibnamefont {Simmons}},
  \bibinfo {author} {\bibfnamefont {J.~R.}\ \bibnamefont {Prance}}, \bibinfo
  {author} {\bibfnamefont {J.~K.}\ \bibnamefont {Gamble}}, \bibinfo {author}
  {\bibfnamefont {M.}~\bibnamefont {Friesen}}, \bibinfo {author} {\bibfnamefont
  {D.~E.}\ \bibnamefont {Savage}}, \bibinfo {author} {\bibfnamefont {M.~G.}\
  \bibnamefont {Lagally}}, \bibinfo {author} {\bibfnamefont {S.~N.}\
  \bibnamefont {Coppersmith}}, \ and\ \bibinfo {author} {\bibfnamefont {M.~A.}\
  \bibnamefont {Eriksson}},\ }\href@noop {} {\bibfield  {journal} {\bibinfo
  {journal} {Appl. Phys. Lett.}\ }\textbf {\bibinfo {volume} {99}},\ \bibinfo
  {pages} {233108} (\bibinfo {year} {2011})}\BibitemShut {NoStop}%
\bibitem [{\citenamefont {Prance}\ \emph {et~al.}(2012)\citenamefont {Prance},
  \citenamefont {Shi}, \citenamefont {Simmons}, \citenamefont {Savage},
  \citenamefont {Lagally}, \citenamefont {Schreiber}, \citenamefont
  {Vandersypen}, \citenamefont {Friesen}, \citenamefont {Joynt}, \citenamefont
  {Coppersmith},\ and\ \citenamefont {Eriksson}}]{Prance2012}%
  \BibitemOpen
  \bibfield  {author} {\bibinfo {author} {\bibfnamefont {J.~R.}\ \bibnamefont
  {Prance}}, \bibinfo {author} {\bibfnamefont {Z.}~\bibnamefont {Shi}},
  \bibinfo {author} {\bibfnamefont {C.~B.}\ \bibnamefont {Simmons}}, \bibinfo
  {author} {\bibfnamefont {D.~E.}\ \bibnamefont {Savage}}, \bibinfo {author}
  {\bibfnamefont {M.~G.}\ \bibnamefont {Lagally}}, \bibinfo {author}
  {\bibfnamefont {L.~R.}\ \bibnamefont {Schreiber}}, \bibinfo {author}
  {\bibfnamefont {L.~M.~K.}\ \bibnamefont {Vandersypen}}, \bibinfo {author}
  {\bibfnamefont {M.}~\bibnamefont {Friesen}}, \bibinfo {author} {\bibfnamefont
  {R.}~\bibnamefont {Joynt}}, \bibinfo {author} {\bibfnamefont {S.~N.}\
  \bibnamefont {Coppersmith}}, \ and\ \bibinfo {author} {\bibfnamefont {M.~A.}\
  \bibnamefont {Eriksson}},\ }\href@noop {} {\bibfield  {journal} {\bibinfo
  {journal} {Phys. Rev. Lett.}\ }\textbf {\bibinfo {volume} {108}},\ \bibinfo
  {pages} {046808} (\bibinfo {year} {2012})}\BibitemShut {NoStop}%
\bibitem [{\citenamefont {Shaji}\ \emph {et~al.}(2008)\citenamefont {Shaji},
  \citenamefont {Simmons}, \citenamefont {Thalakulam}, \citenamefont {Klein},
  \citenamefont {Qin}, \citenamefont {Luo}, \citenamefont {Savage},
  \citenamefont {Lagally}, \citenamefont {Rimberg}, \citenamefont {Joynt},
  \citenamefont {Friesen}, \citenamefont {Blick}, \citenamefont {Coppersmith},\
  and\ \citenamefont {Eriksson}}]{Shaji2008}%
  \BibitemOpen
  \bibfield  {author} {\bibinfo {author} {\bibfnamefont {N.}~\bibnamefont
  {Shaji}}, \bibinfo {author} {\bibfnamefont {C.~B.}\ \bibnamefont {Simmons}},
  \bibinfo {author} {\bibfnamefont {M.}~\bibnamefont {Thalakulam}}, \bibinfo
  {author} {\bibfnamefont {L.~J.}\ \bibnamefont {Klein}}, \bibinfo {author}
  {\bibfnamefont {H.}~\bibnamefont {Qin}}, \bibinfo {author} {\bibfnamefont
  {H.}~\bibnamefont {Luo}}, \bibinfo {author} {\bibfnamefont {D.~E.}\
  \bibnamefont {Savage}}, \bibinfo {author} {\bibfnamefont {M.~G.}\
  \bibnamefont {Lagally}}, \bibinfo {author} {\bibfnamefont {A.~J.}\
  \bibnamefont {Rimberg}}, \bibinfo {author} {\bibfnamefont {R.}~\bibnamefont
  {Joynt}}, \bibinfo {author} {\bibfnamefont {M.}~\bibnamefont {Friesen}},
  \bibinfo {author} {\bibfnamefont {R.~H.}\ \bibnamefont {Blick}}, \bibinfo
  {author} {\bibfnamefont {S.~N.}\ \bibnamefont {Coppersmith}}, \ and\ \bibinfo
  {author} {\bibfnamefont {M.~A.}\ \bibnamefont {Eriksson}},\ }\href@noop {}
  {\bibfield  {journal} {\bibinfo  {journal} {Nature Physics}\ }\textbf
  {\bibinfo {volume} {4}},\ \bibinfo {pages} {540} (\bibinfo {year}
  {2008})}\BibitemShut {NoStop}%
\bibitem [{\citenamefont {Lu}\ \emph {et~al.}(2009)\citenamefont {Lu},
  \citenamefont {Tsui}, \citenamefont {Lee},\ and\ \citenamefont
  {Liu}}]{Lu2009}%
  \BibitemOpen
  \bibfield  {author} {\bibinfo {author} {\bibfnamefont {T.~M.}\ \bibnamefont
  {Lu}}, \bibinfo {author} {\bibfnamefont {D.~C.}\ \bibnamefont {Tsui}},
  \bibinfo {author} {\bibfnamefont {C.-H.}\ \bibnamefont {Lee}}, \ and\
  \bibinfo {author} {\bibfnamefont {C.~W.}\ \bibnamefont {Liu}},\ }\href@noop
  {} {\bibfield  {journal} {\bibinfo  {journal} {Appl. Phys. Lett.}\ }\textbf
  {\bibinfo {volume} {94}},\ \bibinfo {pages} {182102} (\bibinfo {year}
  {2009})}\BibitemShut {NoStop}%
\bibitem [{\citenamefont {Lu}\ \emph {et~al.}(2011)\citenamefont {Lu},
  \citenamefont {Bishop}, \citenamefont {Pluym}, \citenamefont {Means},
  \citenamefont {Kotula}, \citenamefont {Cederberg}, \citenamefont {Tracy},
  \citenamefont {Dominguez}, \citenamefont {Lilly},\ and\ \citenamefont
  {Carroll}}]{Lu2011}%
  \BibitemOpen
  \bibfield  {author} {\bibinfo {author} {\bibfnamefont {T.~M.}\ \bibnamefont
  {Lu}}, \bibinfo {author} {\bibfnamefont {N.~C.}\ \bibnamefont {Bishop}},
  \bibinfo {author} {\bibfnamefont {T.}~\bibnamefont {Pluym}}, \bibinfo
  {author} {\bibfnamefont {J.}~\bibnamefont {Means}}, \bibinfo {author}
  {\bibfnamefont {P.~G.}\ \bibnamefont {Kotula}}, \bibinfo {author}
  {\bibfnamefont {J.}~\bibnamefont {Cederberg}}, \bibinfo {author}
  {\bibfnamefont {L.~A.}\ \bibnamefont {Tracy}}, \bibinfo {author}
  {\bibfnamefont {J.}~\bibnamefont {Dominguez}}, \bibinfo {author}
  {\bibfnamefont {M.~P.}\ \bibnamefont {Lilly}}, \ and\ \bibinfo {author}
  {\bibfnamefont {M.~S.}\ \bibnamefont {Carroll}},\ }\href@noop {} {\bibfield
  {journal} {\bibinfo  {journal} {Appl. Phys. Lett.}\ }\textbf {\bibinfo
  {volume} {99}},\ \bibinfo {pages} {043101} (\bibinfo {year}
  {2011})}\BibitemShut {NoStop}%
\bibitem [{\citenamefont {Xiao}, \citenamefont {House},\ and\ \citenamefont
  {Jiang}(2010{\natexlab{b}})}]{Xiao2010APL}%
  \BibitemOpen
  \bibfield  {author} {\bibinfo {author} {\bibfnamefont {M.}~\bibnamefont
  {Xiao}}, \bibinfo {author} {\bibfnamefont {M.~G.}\ \bibnamefont {House}}, \
  and\ \bibinfo {author} {\bibfnamefont {H.~W.}\ \bibnamefont {Jiang}},\
  }\href@noop {} {\bibfield  {journal} {\bibinfo  {journal} {Appl. Phys.
  Lett.}\ }\textbf {\bibinfo {volume} {97}},\ \bibinfo {pages} {032103}
  (\bibinfo {year} {2010}{\natexlab{b}})}\BibitemShut {NoStop}%
\bibitem [{\citenamefont {Borselli}\ \emph
  {et~al.}(2011{\natexlab{a}})\citenamefont {Borselli}, \citenamefont {Eng},
  \citenamefont {Croke}, \citenamefont {Maune}, \citenamefont {Huang},
  \citenamefont {Ross}, \citenamefont {Kiselev}, \citenamefont {Deelman},
  \citenamefont {Alvarado-Rodriguez}, \citenamefont {Schmitz}, \citenamefont
  {Sokolich}, \citenamefont {Holabird}, \citenamefont {Hazard}, \citenamefont
  {Gyure},\ and\ \citenamefont {Hunter}}]{Borselli_FG}%
  \BibitemOpen
  \bibfield  {author} {\bibinfo {author} {\bibfnamefont {M.~G.}\ \bibnamefont
  {Borselli}}, \bibinfo {author} {\bibfnamefont {K.}~\bibnamefont {Eng}},
  \bibinfo {author} {\bibfnamefont {E.~T.}\ \bibnamefont {Croke}}, \bibinfo
  {author} {\bibfnamefont {B.~M.}\ \bibnamefont {Maune}}, \bibinfo {author}
  {\bibfnamefont {B.}~\bibnamefont {Huang}}, \bibinfo {author} {\bibfnamefont
  {R.~S.}\ \bibnamefont {Ross}}, \bibinfo {author} {\bibfnamefont {A.~A.}\
  \bibnamefont {Kiselev}}, \bibinfo {author} {\bibfnamefont {P.~W.}\
  \bibnamefont {Deelman}}, \bibinfo {author} {\bibfnamefont {I.}~\bibnamefont
  {Alvarado-Rodriguez}}, \bibinfo {author} {\bibfnamefont {A.~E.}\ \bibnamefont
  {Schmitz}}, \bibinfo {author} {\bibfnamefont {M.}~\bibnamefont {Sokolich}},
  \bibinfo {author} {\bibfnamefont {K.~S.}\ \bibnamefont {Holabird}}, \bibinfo
  {author} {\bibfnamefont {T.~M.}\ \bibnamefont {Hazard}}, \bibinfo {author}
  {\bibfnamefont {M.~F.}\ \bibnamefont {Gyure}}, \ and\ \bibinfo {author}
  {\bibfnamefont {A.~T.}\ \bibnamefont {Hunter}},\ }\href {\doibase
  10.1063/1.3623479} {\bibfield  {journal} {\bibinfo  {journal} {Appl. Phys.
  Lett.}\ }\textbf {\bibinfo {volume} {99}},\ \bibinfo {eid} {063109} (\bibinfo
  {year} {2011}{\natexlab{a}})}\BibitemShut {NoStop}%
\bibitem [{\citenamefont {Maune}\ \emph {et~al.}(2012)\citenamefont {Maune},
  \citenamefont {Borselli}, \citenamefont {Huang}, \citenamefont {Ladd},
  \citenamefont {Deelman}, \citenamefont {Holabird}, \citenamefont {Kiselev},
  \citenamefont {Alvarado-Rodriguez}, \citenamefont {Ross}, \citenamefont
  {Schmitz}, \citenamefont {Sokolich}, \citenamefont {Watson}, \citenamefont
  {Gyure},\ and\ \citenamefont {Hunter}}]{Maune2012}%
  \BibitemOpen
  \bibfield  {author} {\bibinfo {author} {\bibfnamefont {B.~M.}\ \bibnamefont
  {Maune}}, \bibinfo {author} {\bibfnamefont {M.~G.}\ \bibnamefont {Borselli}},
  \bibinfo {author} {\bibfnamefont {B.}~\bibnamefont {Huang}}, \bibinfo
  {author} {\bibfnamefont {T.~D.}\ \bibnamefont {Ladd}}, \bibinfo {author}
  {\bibfnamefont {P.~W.}\ \bibnamefont {Deelman}}, \bibinfo {author}
  {\bibfnamefont {K.~S.}\ \bibnamefont {Holabird}}, \bibinfo {author}
  {\bibfnamefont {A.~A.}\ \bibnamefont {Kiselev}}, \bibinfo {author}
  {\bibfnamefont {I.}~\bibnamefont {Alvarado-Rodriguez}}, \bibinfo {author}
  {\bibfnamefont {R.~S.}\ \bibnamefont {Ross}}, \bibinfo {author}
  {\bibfnamefont {A.~E.}\ \bibnamefont {Schmitz}}, \bibinfo {author}
  {\bibfnamefont {M.}~\bibnamefont {Sokolich}}, \bibinfo {author}
  {\bibfnamefont {C.~A.}\ \bibnamefont {Watson}}, \bibinfo {author}
  {\bibfnamefont {M.~F.}\ \bibnamefont {Gyure}}, \ and\ \bibinfo {author}
  {\bibfnamefont {A.~T.}\ \bibnamefont {Hunter}},\ }\href@noop {} {\bibfield
  {journal} {\bibinfo  {journal} {Nature}\ }\textbf {\bibinfo {volume} {481}},\
  \bibinfo {pages} {344} (\bibinfo {year} {2012})}\BibitemShut {NoStop}%
\bibitem [{\citenamefont {Yang}\ \emph {et~al.}(2011)\citenamefont {Yang},
  \citenamefont {Lim}, \citenamefont {Zwanenburg},\ and\ \citenamefont
  {Dzurak}}]{Yang2011}%
  \BibitemOpen
  \bibfield  {author} {\bibinfo {author} {\bibfnamefont {C.~H.}\ \bibnamefont
  {Yang}}, \bibinfo {author} {\bibfnamefont {W.~H.}\ \bibnamefont {Lim}},
  \bibinfo {author} {\bibfnamefont {F.~A.}\ \bibnamefont {Zwanenburg}}, \ and\
  \bibinfo {author} {\bibfnamefont {A.~S.}\ \bibnamefont {Dzurak}},\
  }\href@noop {} {\bibfield  {journal} {\bibinfo  {journal} {AIP Adv.}\
  }\textbf {\bibinfo {volume} {1}},\ \bibinfo {pages} {042111} (\bibinfo {year}
  {2011})}\BibitemShut {NoStop}%
\bibitem [{\citenamefont {Yang}\ \emph {et~al.}(2012)\citenamefont {Yang},
  \citenamefont {Lim}, \citenamefont {Lai}, \citenamefont {Rossi},
  \citenamefont {Morello},\ and\ \citenamefont {Dzurak}}]{Yang2012}%
  \BibitemOpen
  \bibfield  {author} {\bibinfo {author} {\bibfnamefont {C.~H.}\ \bibnamefont
  {Yang}}, \bibinfo {author} {\bibfnamefont {W.~H.}\ \bibnamefont {Lim}},
  \bibinfo {author} {\bibfnamefont {N.~S.}\ \bibnamefont {Lai}}, \bibinfo
  {author} {\bibfnamefont {A.}~\bibnamefont {Rossi}}, \bibinfo {author}
  {\bibfnamefont {A.}~\bibnamefont {Morello}}, \ and\ \bibinfo {author}
  {\bibfnamefont {A.~S.}\ \bibnamefont {Dzurak}},\ }\href@noop {} {\bibfield
  {journal} {\bibinfo  {journal} {Phys. Rev. B}\ }\textbf {\bibinfo {volume}
  {86}},\ \bibinfo {pages} {115319} (\bibinfo {year} {2012})}\BibitemShut
  {NoStop}%
\bibitem [{\citenamefont {Lim}\ \emph {et~al.}(2011)\citenamefont {Lim},
  \citenamefont {Yang}, \citenamefont {Zwanenburg},\ and\ \citenamefont
  {Dzurak}}]{Lim2011}%
  \BibitemOpen
  \bibfield  {author} {\bibinfo {author} {\bibfnamefont {W.~H.}\ \bibnamefont
  {Lim}}, \bibinfo {author} {\bibfnamefont {C.~H.}\ \bibnamefont {Yang}},
  \bibinfo {author} {\bibfnamefont {F.~A.}\ \bibnamefont {Zwanenburg}}, \ and\
  \bibinfo {author} {\bibfnamefont {A.~S.}\ \bibnamefont {Dzurak}},\
  }\href@noop {} {\bibfield  {journal} {\bibinfo  {journal} {Nanotechnology}\
  }\textbf {\bibinfo {volume} {22}},\ \bibinfo {pages} {335704} (\bibinfo
  {year} {2011})}\BibitemShut {NoStop}%
\bibitem [{\citenamefont {Lai}\ \emph {et~al.}(2011)\citenamefont {Lai},
  \citenamefont {Lim}, \citenamefont {Yang}, \citenamefont {Zwanenburg},
  \citenamefont {Coish}, \citenamefont {Qassemi}, \citenamefont {Morello},\
  and\ \citenamefont {Dzurak}}]{Lai2011}%
  \BibitemOpen
  \bibfield  {author} {\bibinfo {author} {\bibfnamefont {N.~S.}\ \bibnamefont
  {Lai}}, \bibinfo {author} {\bibfnamefont {W.~H.}\ \bibnamefont {Lim}},
  \bibinfo {author} {\bibfnamefont {C.~H.}\ \bibnamefont {Yang}}, \bibinfo
  {author} {\bibfnamefont {F.~A.}\ \bibnamefont {Zwanenburg}}, \bibinfo
  {author} {\bibfnamefont {W.~A.}\ \bibnamefont {Coish}}, \bibinfo {author}
  {\bibfnamefont {F.}~\bibnamefont {Qassemi}}, \bibinfo {author} {\bibfnamefont
  {A.}~\bibnamefont {Morello}}, \ and\ \bibinfo {author} {\bibfnamefont
  {A.~S.}\ \bibnamefont {Dzurak}},\ }\href@noop {} {\bibfield  {journal}
  {\bibinfo  {journal} {Sci. Rep.}\ }\textbf {\bibinfo {volume} {1}},\ \bibinfo
  {pages} {110} (\bibinfo {year} {2011})}\BibitemShut {NoStop}%
\bibitem [{\citenamefont {Croke}\ \emph {et~al.}(2010)\citenamefont {Croke},
  \citenamefont {Borselli}, \citenamefont {Gyure}, \citenamefont {Bui},
  \citenamefont {Milosavljevic}, \citenamefont {Ross}, \citenamefont
  {Schmitz},\ and\ \citenamefont {Hunter}}]{Croke2010}%
  \BibitemOpen
  \bibfield  {author} {\bibinfo {author} {\bibfnamefont {E.~T.}\ \bibnamefont
  {Croke}}, \bibinfo {author} {\bibfnamefont {M.~G.}\ \bibnamefont {Borselli}},
  \bibinfo {author} {\bibfnamefont {M.~F.}\ \bibnamefont {Gyure}}, \bibinfo
  {author} {\bibfnamefont {S.~S.}\ \bibnamefont {Bui}}, \bibinfo {author}
  {\bibfnamefont {I.~I.}\ \bibnamefont {Milosavljevic}}, \bibinfo {author}
  {\bibfnamefont {R.~S.}\ \bibnamefont {Ross}}, \bibinfo {author}
  {\bibfnamefont {A.~E.}\ \bibnamefont {Schmitz}}, \ and\ \bibinfo {author}
  {\bibfnamefont {A.~T.}\ \bibnamefont {Hunter}},\ }\href {\doibase
  10.1063/1.3280368} {\bibfield  {journal} {\bibinfo  {journal} {Appl. Phys.
  Lett.}\ }\textbf {\bibinfo {volume} {96}},\ \bibinfo {eid} {042101} (\bibinfo
  {year} {2010})}\BibitemShut {NoStop}%
\bibitem [{\citenamefont {Borselli}\ \emph
  {et~al.}(2011{\natexlab{b}})\citenamefont {Borselli}, \citenamefont {Ross},
  \citenamefont {Kiselev}, \citenamefont {Croke}, \citenamefont {Holabird},
  \citenamefont {Deelman}, \citenamefont {Warren}, \citenamefont
  {Alvarado-Rodriguez}, \citenamefont {Milosavljevic}, \citenamefont {Ku},
  \citenamefont {Wong}, \citenamefont {Schmitz}, \citenamefont {Sokolich},
  \citenamefont {Gyure},\ and\ \citenamefont {Hunter}}]{Borselli_VS}%
  \BibitemOpen
  \bibfield  {author} {\bibinfo {author} {\bibfnamefont {M.~G.}\ \bibnamefont
  {Borselli}}, \bibinfo {author} {\bibfnamefont {R.~S.}\ \bibnamefont {Ross}},
  \bibinfo {author} {\bibfnamefont {A.~A.}\ \bibnamefont {Kiselev}}, \bibinfo
  {author} {\bibfnamefont {E.~T.}\ \bibnamefont {Croke}}, \bibinfo {author}
  {\bibfnamefont {K.~S.}\ \bibnamefont {Holabird}}, \bibinfo {author}
  {\bibfnamefont {P.~W.}\ \bibnamefont {Deelman}}, \bibinfo {author}
  {\bibfnamefont {L.~D.}\ \bibnamefont {Warren}}, \bibinfo {author}
  {\bibfnamefont {I.}~\bibnamefont {Alvarado-Rodriguez}}, \bibinfo {author}
  {\bibfnamefont {I.}~\bibnamefont {Milosavljevic}}, \bibinfo {author}
  {\bibfnamefont {F.~C.}\ \bibnamefont {Ku}}, \bibinfo {author} {\bibfnamefont
  {W.~S.}\ \bibnamefont {Wong}}, \bibinfo {author} {\bibfnamefont {A.~E.}\
  \bibnamefont {Schmitz}}, \bibinfo {author} {\bibfnamefont {M.}~\bibnamefont
  {Sokolich}}, \bibinfo {author} {\bibfnamefont {M.~F.}\ \bibnamefont {Gyure}},
  \ and\ \bibinfo {author} {\bibfnamefont {A.~T.}\ \bibnamefont {Hunter}},\
  }\href@noop {} {\bibfield  {journal} {\bibinfo  {journal} {Appl. Phys.
  Lett.}\ }\textbf {\bibinfo {volume} {98}},\ \bibinfo {pages} {123118}
  (\bibinfo {year} {2011}{\natexlab{b}})}\BibitemShut {NoStop}%
\bibitem [{\citenamefont {Wolf}\ and\ \citenamefont
  {Tauber}(1986)}]{Wolf_and_TauberV1}%
  \BibitemOpen
  \bibfield  {author} {\bibinfo {author} {\bibfnamefont {S.}~\bibnamefont
  {Wolf}}\ and\ \bibinfo {author} {\bibfnamefont {R.~N.}\ \bibnamefont
  {Tauber}},\ }\href@noop {} {\emph {\bibinfo {title} {{Silicon Processing for
  the VLSI Era, Vol. 1: Process Technology}}}}\ (\bibinfo  {publisher} {Lattice
  Press},\ \bibinfo {address} {Sunset Beach, CA},\ \bibinfo {year}
  {1986})\BibitemShut {NoStop}%
\bibitem [{\citenamefont {Barthel}\ \emph {et~al.}(2010)\citenamefont
  {Barthel}, \citenamefont {Kj\ae{}rgaard}, \citenamefont {Medford},
  \citenamefont {Stopa}, \citenamefont {Marcus}, \citenamefont {Hanson},\ and\
  \citenamefont {Gossard}}]{Barthel2010}%
  \BibitemOpen
  \bibfield  {author} {\bibinfo {author} {\bibfnamefont {C.}~\bibnamefont
  {Barthel}}, \bibinfo {author} {\bibfnamefont {M.}~\bibnamefont
  {Kj\ae{}rgaard}}, \bibinfo {author} {\bibfnamefont {J.}~\bibnamefont
  {Medford}}, \bibinfo {author} {\bibfnamefont {M.}~\bibnamefont {Stopa}},
  \bibinfo {author} {\bibfnamefont {C.~M.}\ \bibnamefont {Marcus}}, \bibinfo
  {author} {\bibfnamefont {M.~P.}\ \bibnamefont {Hanson}}, \ and\ \bibinfo
  {author} {\bibfnamefont {A.~C.}\ \bibnamefont {Gossard}},\ }\href@noop {}
  {\bibfield  {journal} {\bibinfo  {journal} {Phys. Rev. B}\ }\textbf {\bibinfo
  {volume} {81}},\ \bibinfo {pages} {161308} (\bibinfo {year}
  {2010})}\BibitemShut {NoStop}%
\bibitem [{\citenamefont {Nordberg}\ \emph {et~al.}(2009)\citenamefont
  {Nordberg}, \citenamefont {Stalford}, \citenamefont {Young}, \citenamefont
  {Eyck}, \citenamefont {Eng}, \citenamefont {Tracy}, \citenamefont {Childs},
  \citenamefont {Wendt}, \citenamefont {Grubbs}, \citenamefont {Stevens},
  \citenamefont {Lilly}, \citenamefont {Eriksson},\ and\ \citenamefont
  {Carroll}}]{Nordberg2009}%
  \BibitemOpen
  \bibfield  {author} {\bibinfo {author} {\bibfnamefont {E.~P.}\ \bibnamefont
  {Nordberg}}, \bibinfo {author} {\bibfnamefont {H.~L.}\ \bibnamefont
  {Stalford}}, \bibinfo {author} {\bibfnamefont {R.}~\bibnamefont {Young}},
  \bibinfo {author} {\bibfnamefont {G.~A.~T.}\ \bibnamefont {Eyck}}, \bibinfo
  {author} {\bibfnamefont {K.}~\bibnamefont {Eng}}, \bibinfo {author}
  {\bibfnamefont {L.~A.}\ \bibnamefont {Tracy}}, \bibinfo {author}
  {\bibfnamefont {K.~D.}\ \bibnamefont {Childs}}, \bibinfo {author}
  {\bibfnamefont {J.~R.}\ \bibnamefont {Wendt}}, \bibinfo {author}
  {\bibfnamefont {R.~K.}\ \bibnamefont {Grubbs}}, \bibinfo {author}
  {\bibfnamefont {J.}~\bibnamefont {Stevens}}, \bibinfo {author} {\bibfnamefont
  {M.~P.}\ \bibnamefont {Lilly}}, \bibinfo {author} {\bibfnamefont {M.~A.}\
  \bibnamefont {Eriksson}}, \ and\ \bibinfo {author} {\bibfnamefont {M.~S.}\
  \bibnamefont {Carroll}},\ }\href@noop {} {\bibfield  {journal} {\bibinfo
  {journal} {Appl. Phys. Lett.}\ }\textbf {\bibinfo {volume} {95}},\ \bibinfo
  {pages} {202102} (\bibinfo {year} {2009})}\BibitemShut {NoStop}%
\bibitem [{\citenamefont {Elzerman}\ \emph {et~al.}(2004)\citenamefont
  {Elzerman}, \citenamefont {Hanson}, \citenamefont {van Beveren},
  \citenamefont {Vandersypen},\ and\ \citenamefont
  {Kouwenhoven}}]{Elzerman2004}%
  \BibitemOpen
  \bibfield  {author} {\bibinfo {author} {\bibfnamefont {J.~M.}\ \bibnamefont
  {Elzerman}}, \bibinfo {author} {\bibfnamefont {R.}~\bibnamefont {Hanson}},
  \bibinfo {author} {\bibfnamefont {L.~H.~W.}\ \bibnamefont {van Beveren}},
  \bibinfo {author} {\bibfnamefont {L.~M.~K.}\ \bibnamefont {Vandersypen}}, \
  and\ \bibinfo {author} {\bibfnamefont {L.~P.}\ \bibnamefont {Kouwenhoven}},\
  }\href@noop {} {\bibfield  {journal} {\bibinfo  {journal} {Appl. Phys.
  Lett.}\ }\textbf {\bibinfo {volume} {84}},\ \bibinfo {pages} {4617} (\bibinfo
  {year} {2004})}\BibitemShut {NoStop}%
\bibitem [{\citenamefont {Hanson}\ \emph {et~al.}(2007)\citenamefont {Hanson},
  \citenamefont {Kouwenhoven}, \citenamefont {Petta}, \citenamefont {Tarucha},\
  and\ \citenamefont {Vandersypen}}]{Hanson2007}%
  \BibitemOpen
  \bibfield  {author} {\bibinfo {author} {\bibfnamefont {R.}~\bibnamefont
  {Hanson}}, \bibinfo {author} {\bibfnamefont {L.~P.}\ \bibnamefont
  {Kouwenhoven}}, \bibinfo {author} {\bibfnamefont {J.~R.}\ \bibnamefont
  {Petta}}, \bibinfo {author} {\bibfnamefont {S.}~\bibnamefont {Tarucha}}, \
  and\ \bibinfo {author} {\bibfnamefont {L.~M.~K.}\ \bibnamefont
  {Vandersypen}},\ }\href@noop {} {\bibfield  {journal} {\bibinfo  {journal}
  {Rev. Mod. Phys.}\ }\textbf {\bibinfo {volume} {79}},\ \bibinfo {pages}
  {1217} (\bibinfo {year} {2007})}\BibitemShut {NoStop}%
\bibitem [{Note1()}]{Note1}%
  \BibitemOpen
  \bibinfo {note} {Independent excited-state-spectroscopy measurements\cite
  {Elzerman2004} and magneto-spectroscopy\cite {Borselli_VS,Yang2012,Shi2011}
  confirmed that this pulse amplitude was large enough to indiscriminately load
  into both ground and excited valley states.}\BibitemShut {Stop}%
\bibitem [{\citenamefont {Thalakulam}\ \emph {et~al.}(2011)\citenamefont
  {Thalakulam}, \citenamefont {Simmons}, \citenamefont {Bael}, \citenamefont
  {Rosemeyer}, \citenamefont {Savage}, \citenamefont {Lagally}, \citenamefont
  {Friesen}, \citenamefont {Coppersmith},\ and\ \citenamefont
  {Eriksson}}]{Thalakulam2011}%
  \BibitemOpen
  \bibfield  {author} {\bibinfo {author} {\bibfnamefont {M.}~\bibnamefont
  {Thalakulam}}, \bibinfo {author} {\bibfnamefont {C.~B.}\ \bibnamefont
  {Simmons}}, \bibinfo {author} {\bibfnamefont {B.~J.~V.}\ \bibnamefont
  {Bael}}, \bibinfo {author} {\bibfnamefont {B.~M.}\ \bibnamefont {Rosemeyer}},
  \bibinfo {author} {\bibfnamefont {D.~E.}\ \bibnamefont {Savage}}, \bibinfo
  {author} {\bibfnamefont {M.~G.}\ \bibnamefont {Lagally}}, \bibinfo {author}
  {\bibfnamefont {M.}~\bibnamefont {Friesen}}, \bibinfo {author} {\bibfnamefont
  {S.~N.}\ \bibnamefont {Coppersmith}}, \ and\ \bibinfo {author} {\bibfnamefont
  {M.~A.}\ \bibnamefont {Eriksson}},\ }\href@noop {} {\bibfield  {journal}
  {\bibinfo  {journal} {Phys. Rev. B}\ }\textbf {\bibinfo {volume} {84}},\
  \bibinfo {pages} {045307} (\bibinfo {year} {2011})}\BibitemShut {NoStop}%
\bibitem [{\citenamefont {Simmons}\ \emph {et~al.}(2009)\citenamefont
  {Simmons}, \citenamefont {Thalakulam}, \citenamefont {Rosemeyer},
  \citenamefont {Bael}, \citenamefont {Sackmann}, \citenamefont {Savage},
  \citenamefont {Lagally}, \citenamefont {Joynt}, \citenamefont {Friesen},
  \citenamefont {Coppersmith},\ and\ \citenamefont {Eriksson}}]{Simmons2009}%
  \BibitemOpen
  \bibfield  {author} {\bibinfo {author} {\bibfnamefont {C.~B.}\ \bibnamefont
  {Simmons}}, \bibinfo {author} {\bibfnamefont {M.}~\bibnamefont {Thalakulam}},
  \bibinfo {author} {\bibfnamefont {B.~M.}\ \bibnamefont {Rosemeyer}}, \bibinfo
  {author} {\bibfnamefont {B.~J.~V.}\ \bibnamefont {Bael}}, \bibinfo {author}
  {\bibfnamefont {E.~K.}\ \bibnamefont {Sackmann}}, \bibinfo {author}
  {\bibfnamefont {D.~E.}\ \bibnamefont {Savage}}, \bibinfo {author}
  {\bibfnamefont {M.~G.}\ \bibnamefont {Lagally}}, \bibinfo {author}
  {\bibfnamefont {R.}~\bibnamefont {Joynt}}, \bibinfo {author} {\bibfnamefont
  {M.}~\bibnamefont {Friesen}}, \bibinfo {author} {\bibfnamefont {S.~N.}\
  \bibnamefont {Coppersmith}}, \ and\ \bibinfo {author} {\bibfnamefont {M.~A.}\
  \bibnamefont {Eriksson}},\ }\href@noop {} {\bibfield  {journal} {\bibinfo
  {journal} {Nano Lett.}\ }\textbf {\bibinfo {volume} {9}},\ \bibinfo {pages}
  {3234} (\bibinfo {year} {2009})}\BibitemShut {NoStop}%
\end{thebibliography}
%

\end{document}